\documentclass[aps,prd,onecolumn]{revtex4}
\usepackage{amssymb}
\usepackage{amsmath}
\usepackage{color}
\usepackage{graphicx}
\usepackage{subfig}
\begin{document}

\title{Instability of the Noncommutative Geometry Inspired Black Hole}
\author{Eric Brown, Robert Mann}
 \affiliation{Department of Physics and Astronomy, University of Waterloo, Waterloo, Ontario N2L 3G1, Canada}

\begin{abstract}
Non-commutative geometries have been proposed as an approach to quantum gravity and
have led to the construction of  non-commutative black holes, whose interior singularities are purportedly eliminated due to quantum effects. Here we find evidence that these black holes are in fact unstable, with infalling matter  near the Cauchy (inner) horizon being subject to an infinite blueshift of the type that has been repeatedly demonstrated for  the Reissner-Nordstr\(\ddot{\text{o}}\)m black hole.  This instability is present even when an ultraviolet cutoff
(induced by anticipated non-commutative geometric effects)
to a field propagating in that spacetime is included. We demonstrate this by following an analogous argument made for Reissner-Nordstr\(\ddot{\text{o}}\)m black holes, and conclude that stability is dependent on the surface gravities \(\kappa_-\) and \(\kappa_+\) of the inner and outer horizons respectively. In general if \(\kappa_- > \kappa_+\), as we show to be the case here, then the stability of the Cauchy horizon becomes highly questionable, contrary to recent claims.
\end{abstract}

\maketitle

The development of a convincing and self-contained theory of quantum gravity is arguably the most important and difficult problem in theoretical physics today. While no such theory has yet achieved this feat, a general tenet thought to be characteristic of quantum gravity is the presence of a natural small-scale cutoff of spacetime; a ``graininess" if you will. This cutoff is expected to arise when approaching the Planckian scale. One of the primary applications expected of quantum gravity is a better understanding of curvature singularities predicted by classical general relativity; such divergent behavior is indicative of a breakdown in the classical theory and it is thought that such extremes will be placated by a proper treatment using quantum gravity. As such, much work has been performed towards understanding black holes in the context of quantum gravity theories. Several metric solutions have been derived using arguments from these theories that are free of the \(r=0\) singularities that plague classical solutions \cite{bh1,bh2,bh3,bh4,bh5}. Indeed, when one imposes a small-scale cutoff one should hardly expect to find a divergence which classically results from trying to describe an infinitely small point in spacetime.

Despite being regular at \(r=0\) however, these quantum gravity black holes generically have more than one horizon. One therefore ought to be wary of their overall stability since it is well known that classical black holes with more than one horizon also exhibit divergent behavior at their inner horizon (the Cauchy horizon). This can be seen to arise from the infinite blueshift of external radiation that occurs on the horizon. The example of prime interest here will be the Reissner-Nordstr\(\ddot{\text{o}}\)m black hole \cite{RNstability,RNstability2,poisson}, though this instability is also found in Kerr and Kerr-Newmann black holes. In this paper we investigate whether or not this instability is present in the black hole solution inspired by noncommutative geometry \cite{bh4,bh5}.

One proposed solution to quantum gravity, motivated by the principle of small-scale cutoff, is to impose a noncommutative structure to the spacetime manifold \cite{madore}. The mathematics of noncommutative geometry is a generalization of differential geometry and its application to spacetime results in an uncertainty relation of the form \(\Delta x^\mu \Delta x^\nu \geq \frac{1}{2}|\theta^{\mu \nu}|\), where \(\theta^{\mu \nu}\) is a matrix that quantifies the magnitude of the noncommutative deviation from a classical spacetime manifold; its entries are postulated to be on the order of the Planck area. This uncertainty relation implies that in this formulation there is no notion of a point-like event in spacetime, in the same way that in quantum mechanics there is no notion of an exact point in phase space. In addition to entirely changing the notion of what a manifold is, such a small-scale cutoff also changes the form of quantum fields propagating on the manifold by imposing a natural momentum cutoff on the order of \(1/\sqrt{\theta}\), where \(\theta\) is defined to be the mean value of the matrix entries. Both of these modifications are important in predicting the phenomena arising from this theory. As with other quantum gravity theories, work has progressed in deriving black hole solutions inspired by noncommutative geometry \cite{bh4,bh5}. The solution corresponding to the classical Schwarzschild metric is
\begin{align} \label{metric}
    ds^2&=-f(r)dt^2+f(r)^{-1}dr^2+r^2d\Omega^2 \nonumber  \\
    f(r)&=1-\frac{2m(r)}{r}  \\
    m(r)&=\frac{2M}{\sqrt{\pi}}\gamma\left(\frac{3}{2},\frac{r^2}{4\theta}\right) \nonumber
\end{align}
The mass function \(m(r)\) is proportional to the total mass \(M\) and a lower, incomplete gamma function \(\gamma(a,b)=\int_0^b x^{a-1}e^{-x}dx\). In the classical limit \(\theta \rightarrow 0\) we see that \(m(r) \rightarrow M\) and Eq. (\ref{metric}) reverts to the Schwarzschild metric.

\begin{figure}[h]
  \centering
  \subfloat[\(M=1.9\sqrt{\theta}\)]{\label{metfunc1}\includegraphics[width=0.3\textwidth]{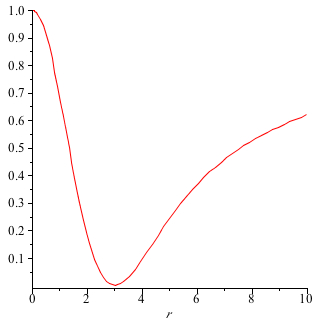}}
  \;\;\;\;\;\;\;\;\;\;\;\;\;\;\;\;
  \subfloat[\(M=4\sqrt{\theta}\)]{\label{metfunc2}\includegraphics[width=0.3\textwidth]{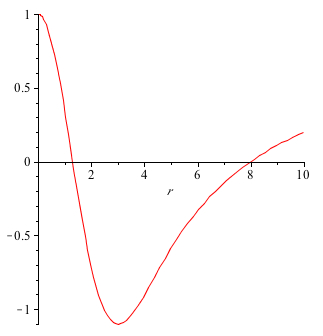}}
  \caption{Metric function \(f(r)\)}
  \label{metfunc}
\end{figure}

This metric admits two horizons provided \(M \gtrsim 1.9\sqrt{\theta}\), below which point it becomes extremal. An example of this is shown in Fig. (\ref{metfunc}) where we plot the metric function \(f(r)\) with \(\theta=1\) and the two cases \(M=1.9\) and \(M=4\). The limiting value of approximately \(1.9\) is determined from the conditions \(f(r_0)=f'(r_0)=0\) for some value \(r_0\) (the radius of the extremal horizon). The condition \(f'(r_0)=0\) can be used to find \(r_0\); one can show that it is given by \(r_0=2\sqrt{\theta x}\) where \(x\) solves the equation \(\gamma(3/2,x)=2x^{3/2}e^{-x}\). The value \(x\) is numerically determined to be \(x \approx 2.284\), with which the other condition \(f(r_0)=0\) can be used to obtain \(M/\sqrt{\theta}=\sqrt{\pi}x^{-1}e^x/4 \approx 1.904\). We will only consider conditions under which two horizons are present, and we will label the outer and inner horizons \(r_+\) and \(r_-\) respectively.

A preliminary investigation into the stability of this black hole claimed that it was stable with respect to blueshifted radiation (with an ultraviolet cutoff) at the Cauchy horizon, since the energy density of an ingoing field pulse (specifically, a Gaussian pulse since a \(\delta\)-function pulse is forbidden by the ultraviolet cutoff) was found to remain finite upon reaching the Cauchy horizon \cite{nic}. However this investigation neglected to examine perturbative scatterings from the primary pulse  that occurs near the horizons. We will show here that such scattered waves to first order do have diverging energy densities at the Cauchy horizon, a result which, while not conclusive, is highly indicative of overall instability. Such a calculation was performed in \cite{RNstability} to indicate the instability of the Reissner-Nordstr\(\ddot{\text{o}}\)m black hole, and our analysis here will follow from the same line of reasoning.

We will find it convenient to use the tortoise coordinate \(r^*\) and null coordinates \(u\) and \(v\), defined as
\begin{align}
    dr^* &\equiv f(r)^{-1}dr \\
    u\equiv r^*-t \, , \;\;\;&\;\;\;\;\;\;\;\;\;\;\; v\equiv r^*+t
\end{align}

The non-commutative structure of spacetime suggests that there is an ultraviolet cutoff to matter fields, and consistent with previous work  \cite{nic} we shall impose the same on a massless matter field \(\phi\) by including a factor of \(e^{-\omega^2 \theta/4}\) in the mode expansion. Neglecting angular dependence, this modified field takes the form
\begin{align} \label{field}
    \phi(r^*,t)=\int_{-\infty}^\infty \frac{d\omega}{r} e^{-\omega^2 \theta/4}e^{-i\omega t}\psi_\omega
\end{align}
where \(\psi_\omega\) solves
\begin{align}  \label{simply}
    \frac{d^2 \psi_\omega}{dr^{*2}}+\left(\omega^2-V(r^*)\right)\psi_\omega=0
\end{align}
and the potential is given by
\begin{align}
    V(r^*(r))=\frac{f(r)}{r}\left(\frac{\ell(\ell+1)}{r}+f'(r)\right)
\end{align}
where \(\ell\) represents the fields angular mode. \(V(r^*)\) vanishes at the horizons and near them can be shown to decay as
\begin{align}
     V(r^*) &\propto e^{2 \kappa_+ r^*} \, , \,\,\, {\rm for} \,\,\,  r \rightarrow r_+ \,\,\, {\rm or} \,\,\, r^* \rightarrow - \infty \nonumber \\
     V(r^*) &\propto e^{- 2 \kappa_- r^*} \, , \,\,\, {\rm for} \,\,\,  r \rightarrow r_- \,\,\, {\rm or} \,\,\, r^* \rightarrow + \infty
\end{align}
where \(\kappa_\pm\) are the surface gravities at the two horizons, defined as
\begin{align}
    \kappa_\pm \equiv \pm \frac{1}{2} f'(r_\pm)
\end{align}

We are principally interested in the field near the inner horizon \(r_-\) since this is where stability comes into question. \(\psi_\omega\) will in general consist of two linearly independent solutions corresponding to left-moving (ingoing) and right-moving (outgoing) waves traveling along surfaces of constant \(v\) and \(u\) respectively. Thus, at the inner horizon the field will be of the form
\begin{align} \label{total}
    \phi \Big |_{r_-} \sim \frac{1}{r_-}\left[g^{(-)}(u)+g^{(+)}(v)\right]
\end{align}
We will need to derive the form of \(g^{(-)}\) and \(g^{(+)}\) eventually. Assuming that they are given, however, we can compute the energy density \(\rho\) of the field as measured by a freely falling observer near the horizon with four-velocity \(U^\alpha\); we have \(\rho=\phi_{, \alpha} \phi_{, \beta} U^\alpha U^\beta+\frac{1}{2}\phi_{, \alpha} \phi^{,\alpha}\). However, since \(u,v=\text{const}\) are null surfaces, the form of \(\phi\) near the horizons implies that this will be dominated by the \(|\phi_{,\alpha}U^\alpha|^2\) term.

The four-velocity of a timelike, radial geodesic is found to be
\begin{align}
    U^t=\frac{E}{f}\, , \;\;\;\;\;\;\;\;\; U^r=-\sqrt{E^2-f}
\end{align}
where we define \(U^r\) to be negative since this will always be the case between the horizons \(r_- < r < r_+\) (since \(r\) necessarily decreases for any observer in this region). \(E\) is a constant of the motion, the sign of which gives the direction of travel between the horizons; \(E>0\) corresponds to a left-moving observer and \(E<0\) to a right-moving one.

Our goal is to compute the energy density \(\rho \propto |U^\alpha g^{(\pm)}_{,\alpha}|^2\). It is easily seen that \(g^{(\pm)}_{,t}=\pm g^{(\pm)'}\) and \(g^{(\pm)}_{,r}=f^{-1}g^{(\pm)'}\), where \(g^{(\pm)'}\) is the derivative of \(g^{(\pm)}\) with respect to \(v\) or \(u\) depending on the sign. With this we obtain
\begin{align}
    U^\alpha g^{(\pm)}_{,\alpha}&=\frac{g^{(\pm)'}}{f}\left(\pm E - |E^2-f|^{1/2}\right)
\end{align}

Let us examine this result in the limit \(r\rightarrow r_-\). Recalling that \(f\) vanishes at the horizons, for a left-moving observer (\(E>0\)) we see that \(U^\alpha g^{(+)}_{,\alpha}\) remains finite but \(U^\alpha g^{(-)}_{,\alpha}\) diverges, unless of course \(g^{(-)'}\) can compensate for the divergence. Conversely, for a right-moving observer (\(E<0\)) \(U^\alpha g^{(+)}_{,\alpha}\) diverges while \(U^\alpha g^{(-)}_{,\alpha}\) remains finite. Assuming the black hole is non-extremal the metric function near \(r_-\) is \(f\simeq -2\kappa_- (r-r_-)\), and so the divergence goes as \(U^\alpha g^{(\pm)}_{,\alpha} \propto g^{(\pm)'}(r-r_-)^{-1}\) for \(r \rightarrow r_-\).

From the form of \(f\) near \(r_-\) we also obtain the tortoise coordinate: \(dr^* \simeq \frac{-dr}{2\kappa_- (r-r_-)}\), thus
\begin{align}
    r^* \simeq -(2\kappa_-)^{-1}\ln |r-r_-|
\end{align}

For a right-moving observer near \(r_-\) it is easily shown that \(\frac{dr^*}{dt}=\frac{dr^*}{dr}\frac{U^r}{U^t} \simeq 1\), from which we obtain
\begin{align}
    -v=-t-r^*&\simeq -2r^*+\text{const}\simeq \kappa_-^{-1}\log|r-r_-|+\text{const} \nonumber \\
    \implies \;\;\;\;\;\;\;\; (r-r_-)^{-1} &\propto e^{\kappa_- v}
\end{align}

This gives us the form of the divergence in \(U^\alpha g^{(+)}_{,\alpha}\) expressed in null coordinates (recall that as \(r\rightarrow r_-\), \(v\rightarrow \infty\) for a right-moving observer and \(u \rightarrow \infty\) for a left-moving one). That is,
\begin{align} \label{cond}
    U^\alpha g^{(+)}_{,\alpha} \propto g^{(+)'} e^{\kappa_- v} \;\;\; \text{for} \;\; r\simeq r_-
    \;\; \text{and} \;\; E<0
\end{align}

Thus, if the inner horizon is to remain stable then \(g^{(+)'}\) must decay at least as fast as \(e^{-\kappa_- v}\) in order to placate the divergence as \(v\rightarrow \infty\). A similar analysis shows that \(g^{(-)'}\) must decay at least as fast as \(e^{-\kappa_- u}\) in order to stop the divergence of \(U^\alpha g^{(-)}_{,\alpha}\) as \(u\rightarrow \infty\) for observers with \(E>0\). Our goal now is now to compute these quantities to determine stability.

In order to compute \(g^{(\pm)'}\) we reproduce a calculation used in \cite{RNstability} to determine the inner horizon stability of the Reissner-Nordstr\(\ddot{\text{o}}\)m black hole. Since we are only interested in the field near the horizons, where the potential is exceedingly small, we can decompose the total solution to Eq. (\ref{simply}), which we will call \(\Psi_\omega\), into the zero-potential solution \(\psi_\omega\) plus an infinitesimal perturbation \(\epsilon_\omega\) produced by the small potential: \(\Psi_\omega=\psi_\omega+\epsilon_\omega\). We set the initial conditions to consist only of ingoing waves and we choose the time dependence to be \(e^{-i\omega t}\); this requires that \(\psi_\omega=e^{-i\omega r^*}\). An ingoing (left-moving) wave will scatter off of the small potential near the horizons, and these scatterings are represented by \(\epsilon_\omega\). In the analysis performed in \cite{nic} only the the main waves \(\psi_\omega\) were considered, and it was found that the energy density due to these waves remains finite on \(r_-\). However they neglected to consider scattered waves \(\epsilon_\omega\), and it is these waves that (as we will show) give rise to instability, for reasons similar to that for a Reissner-Nordstr\(\ddot{\text{o}}\)m black hole \cite{RNstability}.

\begin{figure}[h]
 \centering
  \includegraphics[width=0.5\textwidth]{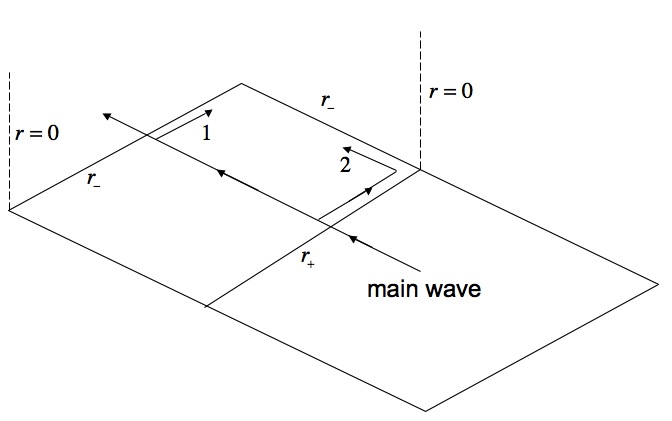}
  \caption{Displaying the different scatterings off of the main wave which are of potential interest.}
\label{scat}
  \end{figure}

There are two scatterings that are of potential interest with regard to stability at the inner horizon. Fig. \ref{scat} shows a section of the Penrose diagram (including the external universe and the \(r_- < r < r_+\) region) with the two scatterings. The first of these consists of right-moving waves traveling along the left branch of \(r_-\); these would have scattered off of the main wave as it neared \(r_-\) and they are labeled with a 1 in Fig. (\ref{scat}). For these waves we must check the form of \(g^{(-)'}\) to test for stability. The second scattering of interest consists of left-moving waves which travel along the right branch of \(r_-\). These waves can form by the following process: consider a scattering produced just after the main wave has entered from the outer horizon; it would be right-moving and traveling along \(r_+\). As this wave approaches the intersection of \(r_+\) and \(r_-\) in the Penrose diagram it enters a region of strong potential. Seeing as the wave magnitude is assumed to be very small, a strong potential would be expected to scatter this wave again with effectively 100\% efficiency; in this case the entire wave is scattered such that it is now a left-moving wave traveling along the right branch of \(r_-\) as labeled by a 2 in Fig. (\ref{scat}). For these waves we must check the form of \(g^{(+)'}\) to test for stability.

In order to solve for \(\epsilon_\omega\) we use a Green's function \(G_{\omega 0}(r^*,y^*)\),
\begin{align} \label{cangreen}
    \left(\frac{\partial^2}{\partial r^{*2}}+\omega^2\right)G_{\omega 0}(r^*,y^*)=\delta (r^*-y^*)
\end{align}
Having chosen an \(e^{-i\omega t}\) time dependence, we opt for the solution
\begin{align} \label{cangreen2}
    G_{\omega 0}(r^*,y^*)=
    \begin{cases}
        \frac{1}{2i\omega}e^{i\omega(r^*-y^*)} & \text{if} \;\; r^* > y^* \\
        \frac{1}{2i\omega}e^{-i\omega(r^*-y^*)} & \text{if} \;\; r^* < y^*
    \end{cases}
\end{align}

Now, since \(\Psi_\omega=\psi_\omega+\epsilon_\omega\) solves Eq. (\ref{simply}) while \(\psi_\omega\) solves the same equation with the potential set to zero, we have that \(\epsilon_\omega\) approximately solves
\begin{align}
    \left(\frac{\partial^2}{\partial r^{*2}}+\omega^2\right)\epsilon_\omega(r^*)=V(r^*)\psi_\omega(r^*)
\end{align}
the solution of which is
\begin{align}
    \epsilon_\omega(r^*)=\int_{-\infty}^\infty G_{\omega 0}(r^*,y^*)V(y^*)\psi_\omega(y^*)dy^*
\end{align}

We must pause now as the form of \(\epsilon_\omega\) is not actually the field that we should use if we are considering a noncommutative spacetime. This is not surprising once one recognizes that the \(\delta\)-function in Eq. (\ref{cangreen}) is not an allowed object given a noncommutative context. We can therefore not use the canonical Green's function \(G_{\omega 0}(r^*,y^*)\) as we have defined it here. We instead must use a modified Green's function \(G_{\omega \theta}(r^*,y^*)\), presented in \cite{nic2}, which is derived in the noncommutative framework. This modified Green's function takes the form
\begin{align}
    G_{\omega\theta}(r^*,y^*)=e^{\theta\frac{\partial^2}{\partial r^{*2}}}G_{\omega 0}(r^*,y^*)
\end{align}
The operator \(e^{\theta\frac{\partial^2}{\partial r^{*2}}}\) acts to smear out point-like objects in order to accommodate the spacetime uncertainty of noncommutative geometry. When applied to a \(\delta\)-function, for example, it outputs a Gaussian of width \(\sim 1/\sqrt{\theta}\).

Given the solution for \(G_{\omega 0}(r^*,y^*)\) in Eq. (\ref{cangreen2}) we can rather easily compute the modified Green's function. For the \(r^* > y^*\) region we have
\begin{align}
    G_{\omega\theta}(r^*,y^*)&=\left(\sum_{n=0}^{\infty}\frac{1}{n!}\theta^n\frac{\partial^{2n}}{\partial r^{*2n}}\right)\left(\frac{1}{2i\omega}e^{i\omega(r^*-y^*)}\right) \nonumber \\
    &=\frac{1}{2i\omega}e^{i\omega(r^*-y^*)}\sum_{n=0}^\infty \frac{(-1)^n}{n!}(\theta \omega^2)^n \nonumber \\
    &=\frac{1}{2i\omega}e^{i\omega(r^*-y^*)}e^{-\theta \omega^2}
\end{align}
and a similar result follows from the \(r^* < y^*\) region. In total we have
\begin{align}
    G_{\omega\theta}(r^*,y^*)=
    \begin{cases}
        \frac{1}{2i\omega}e^{i\omega(r^*-y^*)}e^{-\theta \omega^2} & \text{if} \;\; r^* > y^* \\
        \frac{1}{2i\omega}e^{-i\omega(r^*-y^*)}e^{-\theta \omega^2} & \text{if} \;\; r^* < y^*
    \end{cases}
\end{align}
The noncommutative-modified perturbation field that we wish to use is then given by
\begin{align}
    \epsilon_{\omega \theta}(r^*)=\int_{-\infty}^\infty G_{\omega\theta}(r^*,y^*)V(y^*)\psi_\omega(y^*)dy^*
\end{align}

Here we are interested in the scattering produced by the outer horizon potential \(V(y^*)=V_0e^{2\kappa_+ y^*}\), \(y^* \rightarrow -\infty\). For purposes of computational ease let us set \(V(y^*)=0\) for \(y^* \geq 0\). We evaluate this integral for \(r^*>0\) since this will be the case when the wave approaches \(r_-\), which is what we are interested in. Identifying \(\psi_\omega=e^{-i\omega r^*}\) we obtain trivially
\begin{align}
    \epsilon_{\omega\theta} (r^*)=\frac{V_0 e^{i\omega r^*}e^{-\theta \omega^2}}{4i\omega(\kappa_+-i\omega)}
\end{align}
Including time dependence we obtain the right-moving wave:
\begin{align}
    e^{-i\omega t}\epsilon_{\omega\theta} (r^*)=\frac{V_0 e^{i\omega u}e^{-\theta \omega^2}}{4i\omega(\kappa_+-i\omega)}
\end{align}

We now need to evaluate the total wave consisting of all modes. Given \(e^{-i\omega t}\psi_\omega=e^{-i\omega v}\) the main wave consists of an ingoing, Gaussian pulse \(\phi=\int \frac{d\omega}{r} e^{-\omega^2 \theta/4}e^{-i\omega t}\psi_\omega=\frac{4\pi}{r}\sqrt{\frac{\pi}{\theta}}e^{-v^2/\theta}\). The total perturbation over all modes is
\begin{align}
    \epsilon_\theta=\frac{1}{r}\int_{-\infty}^\infty e^{-\omega^2 \theta/4}e^{-i\omega t}\epsilon_{\omega\theta} d\omega
    =\frac{1}{r}\int_{-\infty}^\infty \frac{V_0 e^{-5\omega^2 \theta/4} e^{i\omega u}}{4i\omega(\kappa_+-i\omega)} d\omega
\end{align}

As explained above, when this wave approaches the intersection of \(r_+\) and \(r_-\) in the Penrose diagram it will be entering a region of high potential, which can be expected to scatter the small wave with near 100\% efficiency. We assume this condition here, and so the wave after this second scattering will be of the same form as that just derived but traveling leftwards instead of rightwards. Recalling Eq. (\ref{total}), we thus have the form of \(g^{(+)}(v)\) for this wave
\begin{align}
    g^{(+)}(v)=\int_{-\infty}^\infty \frac{V_0 e^{-5\omega^2 \theta/4} e^{-i\omega v}}{4i\omega(\kappa_+-i\omega)} d\omega
\end{align}

In order to test for stability, however, we require the derivative
\begin{align}
   g^{(+)'}(v)=\frac{-V_0}{4}\int_{-\infty}^\infty \frac{e^{-5\omega^2 \theta/4}e^{-i\omega v}}{\kappa_+-i\omega}d\omega
\end{align}
which integrates to
\begin{align}
    g^{(+)'}(v)=\frac{-\pi V_0}{4}e^{5\kappa_+ \theta /4}e^{-\kappa_+ v}
    \operatorname{erfc}\left[\frac{1}{2}\sqrt{5\theta}\left(\kappa_+-\frac{2v}{5\theta}\right)\right]
\end{align}
As \(v \rightarrow \infty\) the complimentary error function approaches \(\operatorname{erfc}\left[\frac{1}{2}\sqrt{5\theta}\left(\kappa_+-\frac{2v}{5\theta}\right)\right] \rightarrow 2\), so it is unimportant in the decay behavior. Therefore, we see that \(g^{(+)'}(v)\) decays as \(e^{-\kappa_+ v}\) for an observer approaching the right branch of \(r_-\). Recall that the necessary condition for stability is that \(g^{(+)'}(v)\) decays at least as fast as \(e^{-\kappa_- v}\), and thus the problem now becomes one of comparing the two surface gravities \(\kappa_\pm\). That is, if \(\kappa_- > \kappa_+\) then \(g^{(+)'}(v)\) does not decay fast enough to suppress the divergence present in Eq. (\ref{cond}) and \(r_-\) is unstable. For the noncommutative  black hole this is indeed the case, as we will show.

First recall that there were two possible divergences that could occur at \(r_-\), the other one generated by fields near the left branch of \(r_-\) and with stability contingent on \(g^{(-)'}(u)\) decaying at least as fast as \(e^{-\kappa_- u}\). In this case a similar analysis can be performed and it can be shown that it \emph{does} decay fast enough, and so the Cauchy horizon is stable in this respect.

Turning to the more interesting case, the surface gravities are defined to be \(\kappa_\pm \equiv \pm \frac{1}{2} f'(r_\pm)\), where the metric function \(f(r)\) is given by Eq. (\ref{metric}). Taking the derivative of \(f\) and using the horizon relation \({2m(r_\pm)}={r_\pm}\) we find
\begin{align}
    \kappa_\pm=\pm \frac{1}{2r_\pm}\left(1-2m'(r_\pm)\right)
\end{align}
The metric function \(m(r)\) is given by a lower, incomplete gamma function, as in Eq. (\ref{metric}). As such, its derivative is easily computed, thus giving us expressions for \(\kappa_\pm\) as functions of \(r_\pm\)
\begin{align}
    m'(r)&=\frac{M}{2\sqrt{\pi}}\frac{r^2}{\theta^{3/2}}e^{-r^2/4\theta} \\
    \implies \;\;\;\;\;\;\; \kappa_\pm&=\pm \frac{1}{2r_\pm}\left(1-\frac{M}{\sqrt{\pi}}
    \frac{r_\pm^2}{\theta^{3/2}}e^{-r_\pm^2/4\theta}\right)  \label{surfgrav}
\end{align}

The task now is to compute the values of \(r_\pm\). Looking at Fig. (\ref{metfunc2}) (plotting the metric function for \(\theta=1\) and \(M=4\)) we see that \(\kappa_->\kappa_+\) (that is, the slope is steeper at \(r_-\) than it is at \(r_+\)), yielding an instability at  \(r_-\) .  To confirm this result in the general  we consider two regimes: \(M \gg \sqrt{\theta}\) and \(M \approx 1.9\sqrt{\theta}\). In the intermediate regime between these values we find that \(\kappa_->\kappa_+\), either numerically or simply by plotting \(f(r)\), as in Fig. (\ref{metfunc2}).

\begin{figure}[h]
  \centering
    \includegraphics[width=0.3\textwidth]{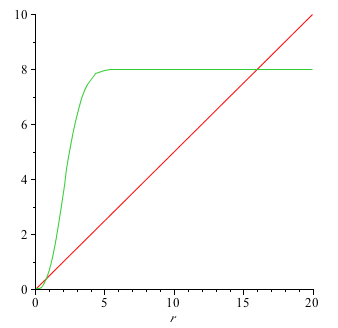}
    \caption{A plot of the mass function \(m(r)\) and the line \(r/2\) for the case of \(\theta=1\) and \(M=8\). The intersections determine the values of \(r_\pm\).}
    \label{inter}
\end{figure}

We first examine the \(M \gg \sqrt{\theta}\) regime. To do so we note that \(r_\pm\) are given by the intersections of the mass function \(m(r)\) and the line \(r/2\); to visualize this we plot the two functions for the case of \(\theta=1\) and \(M=8\) in Fig. (\ref{inter}). We see that in this case \(r_-\) is small and \(r_+\) is equal to \(2M\) to a very good approximation (since the mass function levels off to the constant \(m(r) \simeq M\) by the time \(r/2\) intersects it). As we increase \(M\) with respect to \(\sqrt{\theta}\) these statements only become stronger.

Thus, in this regime we will take \(r_+=2M\) for the outer horizon. For the inner horizon we utilize the fact that \(r_-\) is small and we Taylor expand \(m(r)\) about \(r=0\); it can be shown that the first non-vanishing derivative is \(\frac{d^3m(r)}{dr^3}\Big |_{r=0}=\frac{M}{\sqrt{\pi \theta^3}}\) and so for small \(r\) the mass function is well approximated by a cubic function
\begin{align}
    m(r)=\frac{M}{6\sqrt{\pi \theta^3}}r^3+O(r^5)
\end{align}
We compute \(r_-\) by finding the intersection of this with \(r/2\); along with our result for \(r_+\) we have
\begin{align}
    r_+ \simeq 2M \;\;\; \text{and} \;\;\; r_- \simeq \sqrt{\frac{3\sqrt{\pi \theta^3}}{M}} \;\;\; \text{for} \;\;\; M \gg \sqrt{\theta}
\end{align}
Inserting these results into the surface gravities, Eq. (\ref{surfgrav}), we obtain
\begin{align}
    \kappa_+&=\frac{1}{4M}\left(1-\frac{4}{\sqrt{\pi}}\left(\frac{M}{\sqrt{\theta}}\right)^3
    e^{-M^2/\theta}\right) \\
    \kappa_-&=\frac{1}{2}\sqrt{\frac{M}{3\sqrt{\pi}\theta^{3/2}}}\left(3e^{-3\sqrt{\pi \theta}/4M}-1\right)
\end{align}
from which it can be seen that \(\kappa_+\) will generally decrease and \(\kappa_-\) increase as \(M/\sqrt{\theta}\) increases. Indeed as \(M/\sqrt{\theta} \rightarrow \infty\) we have \(\kappa_+ \rightarrow 1/4M\) and \(\kappa_- \rightarrow \infty\), as should be expected. Having observed that \(\kappa_- > \kappa_+\) for intermediate values of \(M/\sqrt{\theta}\) we thus conclude that this inequality will continue to hold as \(M/\sqrt{\theta}\) increases indefinitely.

\begin{figure}[h]
  \centering
  \subfloat[\(\theta=0.25\)]{\label{kap25}\includegraphics[width=0.5\textwidth]{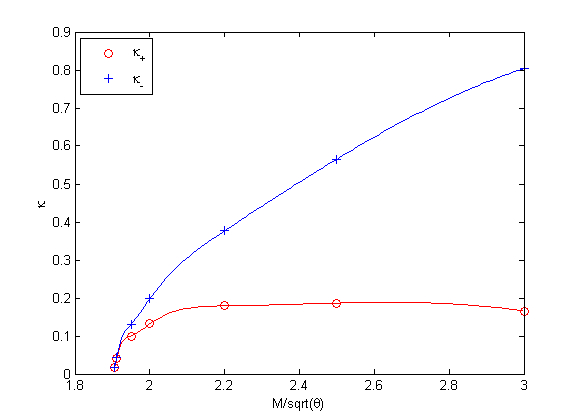}}
  \subfloat[\(\theta=1\)]{\label{kap1}\includegraphics[width=0.5\textwidth]{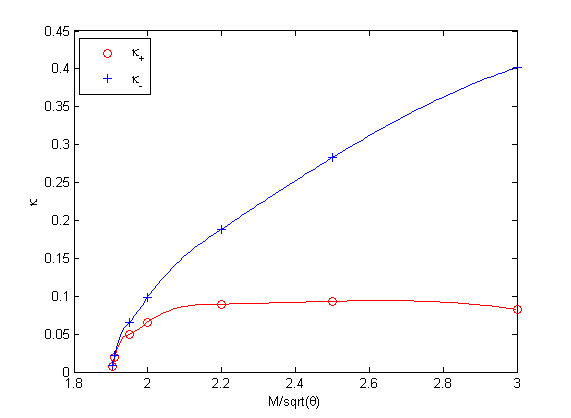}}
  \caption{\(\kappa_\pm\) versus \(M/\sqrt{\theta}\) for two values of \(\theta\). The upper and lower curves correspond to \(\kappa_-\) and \(\kappa_+\) respectively, thus showing that \(\kappa_- > \kappa_+\) for \(M \approx 1.9\sqrt{\theta}\).}
  \label{kap}
\end{figure}

Finally, we wish to examine the \(M \approx 1.9\sqrt{\theta}\) regime; in this case we simply evaluated \(r_\pm\) numerically for several values of \(M\) and \(\theta\) and inserted these values into Eq. (\ref{surfgrav}) to compute \(\kappa_\pm\). The results of this exercise are displayed in Fig. (\ref{kap}) where we plot \(\kappa_\pm\) versus \(M/\sqrt{\theta}\) for \(\theta=0.25\) and \(\theta=1\). As can be seen, \(\kappa_-\) continues to be greater than \(\kappa_+\) as \(M/\sqrt{\theta} \rightarrow 1.9\).

We thus conclude that \(\kappa_- > \kappa_+\) in general, and this implies that the Cauchy horizon of the noncommutative geometry inspired black hole is always unstable with respect to the perturbative scatterings described. Since the scatterings are assumed to be small the fact that they diverge reflects a breakdown in our approximation, and so we can not yet say with absolute certainty that the Cauchy horizon is indeed unstable. However, such a divergence is certainly indicative of instability, casting doubt on previous claims for this black hole \cite{nic}.  We emphasize that this type of analysis is easily performed for any two-horizon black hole, although the scalar field modifications (if any) will be particular to the theory in question. For example, the authors have also applied similar reasoning to a black hole motivated by loop quantum gravity \cite{us}. The results of this work suggest that the loop black hole may have stronger stability properties than the one being scrutinized in this letter.

One might argue that the infalling pulse of scalar field  is not a generic form of infalling matter.
We therefore consider a stress-energy corresponding to an ingoing null dust present on the manifold, but we neglect the back reaction this induces in the metric. This is of course an approximation, but no more of one than was already being assumed in the precious analysis. We will see that the resulting energy density as seen by a timelike observer crossing the Cauchy horizon will, instead of being of the form \(\rho \sim e^{2\kappa_- v}e^{-2\kappa_+ v}\), rather go like \(\rho \sim e^{2\kappa_- v}L(v)\), where \(L(v)\) is the ``luminosity" of the null dust. \(L(v)\) is expected to follow an inverse power law in the large \(v\) limit, as first shown by Price \cite{price}. Thus in this scenario there is no way to avoid diverging energy density as \(v \rightarrow \infty\), unlike in the previous analysis where the competition between two exponentials afforded some hope of this.

First we cast the metric, Eq. (\ref{metric}), in \((v,r,\theta,\phi)\) coordinates; the result is
\begin{align}
    ds^2=-f(r)dv^2+2dvdr+r^2d\Omega^2
\end{align}
where the metric function \(f(r)\) is unchanged.

We then consider a stress-energy tensor which corresponds to an ingoing null dust. It takes the form
\begin{align}
    T_{\alpha \beta}=\mu(r,v)(\partial_\alpha v) (\partial_\beta v)
\end{align}

With these last two equations we determine the form of \(\mu(r,v)\) by computing the conservation condition \(0=T^{\alpha \beta}_{\;\;\; ; \beta}\). The result is that we must have \(\mu(r,v)=A(r)L(v)\), where \(L(v)\) is an arbitrary function of \(v\) and \(A(r)\) must satisfy the equation
\begin{align}
    0=r\frac{dA(r)}{dr}+2A(r)
\end{align}
the solution of which is
\begin{align}
    A(r) \propto \frac{1}{r^2}
\end{align}
Thus, without loss of generality (since \(L(v)\) is arbitrary) we conclude that \(\mu(r,v)\) is given by
\begin{align}
    \mu(r,v)=\frac{1}{4\pi r^2}L(v)
\end{align}

Far from the black hole the spacetime is well approximated by a Schwarzschild solution, and we note that this form of \(\mu(r,v)\) is exactly the same result as that obtained for the Schwarzschild black hole. We thus identify \(L(v)\) with the usual luminosity, and we expect the analysis by Price \cite{price} to hold to a good approximation in our case as well. That is, we have \(L(v) \sim v^{1-p}\) with \(p \geq 12\).

In order to determine the energy density a timelike observer would measure from this dust we also need to specify the four-velocity of the observer. This was previously done for \(U^t\) and \(U^r\), and it is easy to show that \(U^v\) takes the form
\begin{align}
    U^v=\frac{1}{f(r)}\left(E-\sqrt{E^2-f(r)}\right)
\end{align}
where we recall that \(E<0\) for an outgoing observer between the horizons; this is the type of observer that we are interested in.

The energy density measure by such an observer is given by
\begin{align}
    \rho=T_{\alpha \beta}U^\alpha U^\beta=\mu(r,v)(U^v)^2
\end{align}
As the Cauchy horizon is approached we have \(f(r) \rightarrow 0\), and so the negativity of \(E\) implies that \((U^v)^2 \simeq 4E^2/f(r)^2\) as this approach takes place. We also know that near the Cauchy horizon the metric function goes like \(f(r) \sim (r-r_-) \sim e^{-\kappa_- v}\), and so we obtain that the energy density has the behavior
\begin{align}
    \rho \sim e^{2\kappa_- v}L(v)
\end{align}
near the Cauchy horizon. Since \(L(v)\) follows an inverse power law, we conclude that the observer measures diverging energy density upon crossing  the Cauchy horizon.

Of course the preceding null dust calculation is not without its limitations.  Apart from the neglect of back-reaction effects,   it is much less obvious how to include noncommutative modifications to this form of infalling matter, unlike the situation for the scalar field pulse. However the latter is not a generic form of infalling matter.

Notwithstanding such considerations, our calculations provide strong evidence that
 the noncommutative black hole, while retaining finite curvature at \(r=0\), may have an unstable Cauchy horizon.   A more complete treatment of this problem will entail an analysis of the phenomenon of mass inflation \cite{poisson}, in which the mass function of the Reissner-Nordstr\(\ddot{\text{o}}\)m black hole diverges at the Cauchy horizon.   This approach takes into account back-reaction effects from the geometry. The suggestion that mass inflation cannot occur for the noncommutative black hole with ultraviolet restricted fields \cite{nic} appears to be predicated on the claim of finite energy densities, a result that we have now seen does not ensure stability. A more rigorous analysis of mass inflation in this case is needed before one can make conclusions either way.

 \section*{Acknowledgements}
This work was supported by the Natural Sciences and Engineering Research Council of Canada. We are grateful to Piero Nicolini for discussions and helpful correspondence during this work.

\end{document}